\documentclass[10pt, conference, compsocconf]{IEEEtran}

\usepackage[pdftex]{graphicx}

\usepackage[cmex10]{amsmath}
\usepackage{amsfonts}
\usepackage{amssymb}

\begin{document}

\title{Improving files availability for BitTorrent using a diffusion model}

\author{\IEEEauthorblockN{Christian
Napoli\IEEEauthorrefmark{1,*},
Giuseppe Pappalardo \IEEEauthorrefmark{1}, and
Emiliano Tramontana\IEEEauthorrefmark{1}}
\IEEEauthorblockA{\IEEEauthorrefmark{1}Dpt. of Mathematics and
Computer Science, University of Catania, Italy}
an unwanted space
\thanks{*Email: napoli@dmi.unict.it.}
\thanks{2014 IEEE International Conference on Enabling Technologies: Infrastructure for Collaborative Enterprises (WETICE)}}

\markboth{Improving files availability for BitTorrent using a diffusion model -- PREPRINT}%
{Shell \MakeLowercase{\textit{et al.}}: Bare Demo of
IEEEtran.cls for Journals}

 \begin{titlepage}
 \begin{center}
 {\Large \sc PREPRINT VERSION\\}
  \vspace{5mm}
{\huge Improving files availability for BitTorrent using a diffusion model\\}
 \vspace{10mm}
 {\Large C. Napoli, G. Pappalardo, and E. Tramontana\\}
 \vspace{5mm}
{\Large \sc PUBLISHED ON: \bf 2014 IEEE 23rd International WETICE Conference }
 \end{center}
 \vspace{5mm}
 {\Large \sc BIBITEX: \\}
 
@incollection\{Napoli2014Improving\\
year=\{2014\},\\
isbn=\{978-1-4799-4249-7/14\},\\
doi=\{10.1109/WETICE.2014.65\},\\
booktitle=\{23rd International WETICE Conference\},\\
title=\{Improving files availability for BitTorrent using a diffusion model\}, \\
publisher=\{IEEE\},\\
author=\{Napoli, Christian and Pappalardo, Giuseppe and Tramontana Emiliano\},\\
pages=\{191-196\}\\
\}
 \vspace{5mm}
 \begin{center}
Published version copyright \copyright~2014 IEEE \\
\vspace{5mm}
UPLOADED UNDER SELF-ARCHIVING POLICIES\\
NO COPYRIGHT INFRINGEMENT INTENDED \\
 \end{center}
\end{titlepage}

\title{Improving files availability for BitTorrent using a diffusion model}

\author{\IEEEauthorblockN{Christian Napoli, Giuseppe Pappalardo, and
    Emiliano Tramontana} 
\IEEEauthorblockA{Dipartimento di Matematica e Informatica, University of Catania\\
Viale A. Doria 6, 95125 Catania, Italy\\
Email: \{napoli, pappalardo, tramontana\}@dmi.unict.it}}

\maketitle

\begin{abstract}
  The BitTorrent mechanism effectively spreads file fragments by
  copying the rarest fragments first.
  We propose to apply a mathematical model for the diffusion of
  fragments on a P2P in order to take into account both the effects of
  peer distances and the changing availability of peers while time
  goes on.  Moreover, we manage to provide a forecast on the
  availability of a torrent thanks to a neural network that models the
  behaviour of peers on the P2P system.

  The combination of the mathematical model and the neural network
  provides a solution for choosing  file fragments that need to be
  copied first, in order to ensure their continuous availability,
  counteracting possible disconnections by some peers.
\end{abstract}


\section{Introduction}

In peer to peer (P2P) systems using BitTorrent, a shared content
(named torrent) becomes split into fragments and the rarest fragments
are automatically chosen to be sent first to users requesting the
file.  Fragments availability is given by the number of peers storing
a fragment at a given moment, and periodically computed by a server
storing peer ids, fragments held and requested~\cite{Cohen03}.
For computing the priority of fragments to spread, at best
availability has been freshly updated, however peers often leave the
system hence file fragments availability quickly changes, possibly
before the least available fragments have been spread~\cite{Kaune10}.
This occurs so frequently that such a fundamental BitTorrent mechanism
may become ineffective, and as a result some fragments can quickly
become unavailable.
Moreover, when choosing fragments to spread, communication
\emph{latencies}~\cite{repliche08,Novelli07} among peers are not
considered, therefore fragments spreading will surely occur sooner on
peers nearby the one holding the fragment to spread.  As a result, the
furthest peers could disconnect before receiving the whole fragment.

This paper proposes a model for spreading file fragments that
considers both a time-dependent priority for a fragment to spread and
latencies among nodes.  I.e.\ the priority of fragments to spread
gradually changes according to the passing time and fragments
availability.  The  priority variation  is regulated in such a way
that the availability of fragments is maximised while time goes.


Fragments to spread are selected according to the results of our
proposed diffusion model, developed by analogy to a diffusion model on
a porous medium.
Moreover, we propose to characterise the typical availability of a
torrent observed on the P2P system, by using an appropriate neural
network.
Then, the selection of fragments to spread aims at counteracting their
decreasing availability estimated for a later time.
Therefore, the proposed work aims at supporting \emph{Quality of
  Service} and \emph{dependability} of P2P systems by attributing a
priority on both the fragment to spread and the destination peer. This
in turn increases \emph{availability} and
\emph{performances}~\cite{GiuntaMPT12aug,ccpe13,Kaqudai}, as well as
consistency~\cite{BannoMPT10b}.

The rest of the paper is organised as follow. Next section provides
the mathematical representation for the proposed model.
Section~\ref{diffusion} develops the model for the diffusion of
contents on a P2P system.  Section~\ref{wrnn} introduces the neural
network  that predicts the  user behavious. 
Section~\ref{experiments} describes some experiments based on our
proposed model and neural network predictions, as well as the
preliminary results. Section~\ref{related} compares with related work,
and finally conclusions are drawn in Section~\ref{conclusion}.

\section{Mathematical Representation}

In order to put forward our analogy between BitTorrent and a physical
system, some conventions must be chosen and some extrapolations are
needed.  We first describe a continuum system using a continuum
metric, however later on we will single out a few interesting discrete
points of the continuum.  Due to the analogy to BitTorrent, we use a
distance metric (named $\delta$), which will be assimilated to the
network latency among nodes, i.e.\ the hosts on a network holding
seeds, peers or leeches.

For the nodes we use notations $n^i$ or $n^i_\alpha$: the first
indicates a generic $i$-esime node on the BitTorrent network, the second
indicates the $\alpha$-esime node as seen from the $i$-esime node.  Of
course, $n^i_\alpha$ and $n^j_\alpha$ could be different nodes if $i
\neq j$.   Double indexing is needed since when we use something
like $\delta^{ij}$, it will be representing the distance of the
$j$-esime node as measured by the $i$-esime node.  
Moreover, let us express $P^{ij}_k$ as the probability of diffusion of
the $k$-esime file fragment from the $i$-esime node to the $j$-esime
node.  Finally, we distinguish between time and time steps: the first
will be used for a continuum measure of temporal intervals and we will
use for it the latin letter $t$, the second will indicate
computational time steps (e.g.\ the steps of an iterative cycle) and
we will use for it the greek letter $\tau$.  Therefore, while
$\delta^{ij}(t)$ will represent the continuous evolution during time
$t$ of the network latency $\delta$, which is measured by the
$i$-esime node for the distance with the $j$-esime node, the notation
$\delta^{ij}(\tau)$ will represent the latency measured at the
$\tau$-esime step, i.e.\ the time taken by a ping from the $i$-esime
node to the $j$-esime node, only for a specific time step $\tau$.
Finally, we will suppose that each node has the fragment $z_k$ of a
file $z$ and is interested in sharing or obtaining other portions of
the same file, hence we will compute the probability-like function
that expresses how easily the $k$-esime shared fragment is being
copied from the $i$-esime node to the $j$-esime node at a certain step
$\tau$ and we will call it $P^{ij}_k(\tau)$.

Eventually, we are interested in an analytical computation for the
urgency to share a fragment $z_k$ from $n^i$ to $n^j$ during a time
set $\tau$, and we will call it $\chi^{i,j}_k(\tau)$.
In the following sections we will distinguish between an actually
measured value and a value predicted by a neural network using a tilde
for predicted values as in $\tilde x$.

\section{Fragments diffusion on a P2P network}
\label{diffusion}

In our work we compare the file fragments of a shared file to the
diffusion of mass through a porous means.  To embrace this view, it is
mandatory to develop some mathematical tools, which we will explain in
the following.

\subsection{Spaces and metrics}
Users in a P2P BitTorrent network can be represented as elements of a
space where a metric could be given by the corresponding network
communication latency.  Therefore, for each node $n^i \in N$, set of
the nodes, it is possible to define a function
\begin{equation}
\delta:N\times N \rightarrow \mathbb{R} ~~/~~ \delta(n^i,n^j)=
\delta^{ij} ~~~ \forall~ n^i,n^j \in N
\label{eq:delta}
\end{equation}
where $\delta^{ij}$ is the amount of time taken to bring a minimum
amount of data (e.g.\ as for a ping) from $n^i$ to $n^j$.  By using
the given definition of distance, for each node $n^i$, it is also
possible to obtain an ordered list $\Omega^i$ so that
\begin{equation}
\Omega^i = \Big\{ n^i_\alpha \in N \Big\}_{\alpha=0}^{|N|} :
\delta(n^i,n^i_\alpha)\leq\delta(n^i,n^i_{\alpha+1}) 
\label{eq:omega}
\end{equation}
In such a way, the first item of the list will be $n^i_0=n^i$ and the
following items will be ordered according to their network latency as
measured by $n^i$.
Using this complete ordering of peers, it is possible to introduce
the concept of content permeability and diffusion.  Let us consider
the files shared by one user of a P2P system: each file consists of
fragments that can be diffused.  Then, the diffusion of a file
fragment can be analysed in terms of Fick's law.

\subsection{Fick's law and its use for P2P}
Fick's second law is commonly used in physics and chemistry to
describe the change of concentration per  time unit of some element
diffusing into another~\cite{vazquez2006porous}.  
This work proposes an analogy between a P2P system and a physical
system.  The key idea is to model the sharing file fragments as the
diffusion of a substance into a porous means along one dimension.
Different places of the porous means would represent different P2P
nodes, whereas distances along such a one-dimension would be
proportional to the network latencies.
Then, P2P entities would be accommodated into the formalism of
equations~\eqref{eq:delta} and~\eqref{eq:omega}.

Using both the First and Second Fick's laws, the diffusion of a
substance into a means is given as the solution of the following
vectorial differential equation
\begin{equation}
\frac{\partial \Phi}{\partial t} = \nabla \cdot (D \nabla \Phi)
\label{eq:nabla}
\end{equation}
where $\Phi$ is the concentration, $t$ the time and $D$ the
permeability of the means to the diffusing matter. 
%
%
Since this is a separable equation and we use a $1$--dimensional
metric based on the distance $\delta$, and assuming $D$ as constant
among the nodes, equation~\eqref{eq:nabla} can be written as a scalar
differential equation
\begin{equation}
\frac{\partial \Phi}{\partial t} = D \frac{\partial^2 \Phi}{\partial \delta^2} 
\label{eq:partial}
\end{equation}
This partial differential equation, once imposed the initial and
boundaries conditions, admits at least Green's Function as a
particular solution~\cite{bokshtein2005thermodynamics}.
Green's Function lets us study the diffusion dynamics of a single
substance and can be rewritten as solution of the equation~\eqref{eq:partial}
in the form:
\begin{equation}
\Phi(\delta,t)=\Phi_0 \Gamma \Big( \frac{\delta}{ \sqrt{4Dt}}
\Big)~~,~~~~ \Phi_0 = \frac{1}{\sqrt{4 \pi D t}} 
\label{eq:phi}
\end{equation}
where $\Gamma$ is the complementary gaussian error function.

$\Gamma$ function should then be computed by means of a Taylor
expansion. However, to avoid such a computationally difficult task, we
use an approximation proposed in~\cite{Chiani03}, where a pure
exponential approximation for  $\Gamma(x)$ has been obtained,
having an error on the order of $10^{-9}$.  Then, it is possible to
have the following equation
\begin{equation}
\left \{
\begin{array}{rl}
\Phi(\delta,t)\approx&\Phi_0  \left[ \frac{1}{6}
  e^{\left(\Phi_0\delta\right)^2} +
  \frac{1}{2}e^{-\frac{4}{3}\left(\Phi_0\delta\right)^2} \right] \\ 
\\
\Phi_0 =&(4 \pi D t)^{-\frac{1}{2}}
\end{array}
\right .
\label{eq:phisys}
\end{equation}
for every node at a certain distance $\delta \in \mathbb{R}^+$ at a
time $t \in \mathbb{R}^+$.

\subsection{From concentration to probability}

In equation~\eqref{eq:phisys} the scaling factor $\Phi_0$ is a function of the
time $t$.  On the other hand, the used formalism was developed mainly
to focus on the distance $\delta$ and managing $t$ merely as a
parameter.
The above mathematical formalism is valid as long as the distances
$\delta(n^i,n^j)$ remain time-invariant.
The common practice considers the distance between nodes $\delta$ as
time-invariant, however the actual network latencies vary (almost)
continuously, with time, and a stationary $\Omega^i$ ordered set is a
very unlikely approximation for the network.
In our solution, we make 
time-dependent 
the latency embedded into our model.  In turn, this makes it possible
to choose different fragments to be shared as time goes.

%

For the P2P system, the equation~\eqref{eq:phisys} states that a certain file
fragment $z^i_k$ in a node $n^i$ at a time $t_0$ has a probability
$P^{ij}_k(t_0,t)$ to be given (or diffused) to node $n^j$, at a distance
$\delta^{ij}(t_0)$ from $n^i$, within a time $t$, which is
proportional to the $\Phi(\delta, t)$ so that 
\begin{equation}
P^{ij}_k(t_0,t)= p^{ij}_k  \left[ \frac{1}{6}
  e^{\left(p^{ij}_k~\delta^{ij}\right)^2} +
  \frac{1}{2}e^{-\frac{4}{3}\left(p^{ij}_k~\delta^{ij}\right)^2}
\right] \\ 
\label{eq:prob}
\end{equation}
where the function $p^{ij}_k=p^{ij}_k(t_0,t)$ carries both the
diffusion factors and the temporal dynamics. 
And since we are interested in a simple proportion,
not a direct equation, we can also neglect the factor $4\pi$ and then
write $p^{ij}_k$ in the normalised form
\begin{equation}
p^{ij}_k(t,t_0)= \frac{1}{\sqrt{4\pi}} \cdot \frac{1}{\sqrt{D_k(t_0)}} \cdot \frac{1}{\sqrt{t}}
\end{equation}

It is now important to have a proper redefinition of the coefficient
$D$.  Let us say that $T_k$ is the number of users using file fragment
$z_k$ (whether asking or offering it), $S_k$ is the number of seeders
for the file fragment and $\rho_k$ is the mean share ratio of the file
fragment among peers (and leeches), then it is possible to consider
the 'urge' to share the resource as an osmotic pressure which, during
time, varies the coefficient of permeability of the network $D$.
In order to take into account the mutable state in a P2P system, $D$
should vary according to the amounts of available nodes and file
fragments.
We have chosen to define $D$ as
\begin{equation}
D_k (t_0) \triangleq \frac{T_k(t_0)}{S_k(t_0)+\left[T_k(t_0)- S_k(t_0)\right]\rho_k(t_0)}
\end{equation}
by a formal substitution of $D$ with $D_k$ in $\Phi_0$, we obtain the
analytical form of the term $p^{ij}_k$.

\subsection{Discrete time evolution on each node}

Indeed, the physical nature of the adopted law works in the entire
variable space, however for the problem at hand discrete-time
simplifications are needed.  Let us suppose that for a given discrete
time step $\tau=0$ node $n^i$ effectively measures the network
latencies of a set of nodes $\{n^j\}$, then an ordered set $\Omega^i$
as in equation~\eqref{eq:omega} is computed.  Now, every node $n^i$ computes
probability $P^{ij}_k$ for each of its own file fragment $z_k$ and for
every node $n^j$.  This probability corresponds to a statistical
\emph{prevision} of the \emph{possible} file fragments spreading onto
other nodes.

Suppose that for a while no more measures for $\delta$ have been
taken, and at a later discrete time step $\tau$ file fragment $z^i_k$
be copied to the first node to be served,
%
which is chosen according to the minimum probability
of diffusion, latencies and time since last measures were taken (see
following subsection and equation~\eqref{eq:min}).

Moreover, such a file fragment is reaching other nodes if the latency
for such nodes is less than
time $t^i$, computed as
\begin{equation}
t^i_k(\tau)=\sum\limits_{\alpha_k=0}^\tau \delta(n^i,n^i_\alpha)
\label{eq:tik} 
\end{equation}
Index $k$ is used in equation~\eqref{eq:tik} to refer to file fragment
$z^i_k$.

Indeed, since nodes need and offer their own file fragments, the
ordered set of nodes referred by a given node depend on resource
$z_k$, i.e.\ $\Omega^i_k= \{n^i_{\alpha_k}\}$.

It is now possible to have a complete mapping of the probability of
diffusion by reducing the time dependence from $(t_0,t)$ to a single
variable dependence from the discrete time-step $\tau$.  For each
resource $z_k$ as $P^{ij}_k(\tau)$ stated that it is possible to
reduce  $D_k(t_0,t)$ to a one-variable function $D_k(\tau)$ by
assuming that at $t_0$ we have $\tau=0$ and considering only the
values of $D_k(t_0,t)$ when $t$ is the execution moment of a
computational step $\tau$.

\subsection{Assigning priorities and corrections}

Once all the $P^{ij}_k(\tau)$ have been computed, and values stored to
a proper data structure, it is actually simple to determine the most
urgent file fragment to share, which is the resource that has the
least probability to be spread, i.e.\ the $k$ for which
$P^{ij}_k(\tau)$ is minimum.

Furthermore, we consider that while time goes on an old measured
$\delta$  differs from the actual value, hence the measure 
becomes less reliable.
To take into account the staleness of $\delta$ values, we gradually
consider less bound to $\delta$ the choice of a fragment, and this
behaviour is provided by the negative exponential in equation~\eqref{eq:expneg}.
Given enough time, the choice will be based only on the number of
available fragments. However, we consider that by that time a new value for
$\delta$ would have been measured and incorporated again into the
model choosing the fragment.

%

Generally, for nodes having the highest latencies with respect to a
given node $n^i$, more time will be needed to receive a fragment from
the node $n^i$.
We aim at compensating such a delay by incorporating into our model
the inescapable latencies of a P2P network. Therefore, the node that
will receive  a fragment first will be chosen according to its
distance.


In order to model the fact that distant nodes, having the highest
values for $\delta$, will take more time to send or receive file
fragments, we have chosen a decay law. Now it is possible to obtain a
complete time-variant analytical form
of the spreading of file fragments

\begin{equation}
\chi^{ij}_k(\tau) =  \frac{e^{-c \tau \delta^{ij}}}{P^{ij}_k(\tau)}
\label{eq:expneg}
\end{equation}
where the decay constant $c$ can be chosen heuristically, without
harming the said law, and tuned according to other parameters.
If $k$ indicates a file fragment and
$k^*$ the index of the most urgent file fragment to share, this latter is
trivially found as the solution of a maximum problem so that 
\begin{equation}
k^* : \chi^{ij}_{k^*}(\tau) = \max\limits_{k} \left\{ \chi^{ij}_k(\tau) \right\}
\label{eq:min}
\end{equation}
Of course, all the priorities depend on the value of the bi-dimensional
matrix of values of $P^{ij}_k$ (we mark that the index $i$ does not
variate within the same node $n^i$). Among these values,
there is no need to compute elements where $j=i$ and for those
elements where the node $n^j$ is not in the queue for resource
$z_k$.  In both these  cases it is assumed  $P^{ij}_k = 1$.
Moreover, after $n^i$ having completed to transfer $z_k$ to the node
$n^j$, the element of indexes $(j,k)$ is set to 1.  In a similar
fashion, each peer is able to identify a possible resource to ask for
in order to maximize the diffusion of rare ones instead of common
ones.

\section{WRNN predictors and Users behavior}
\label{wrnn}



In order to make the P2P system able to properly react to peaks of
requests, as well as very fast changes of fragments availability
and/or share ratio, we propose an innovative solution based on
\emph{Wavelet Recurrent Neural Networks} (WRNN) to characterise the
user behaviour and producing a short-term forecast.
For a given torrent, the \emph{wavelet analysis} provides compression
and denoising on the observed time series of the amount of users
prividing or requesting fragments; a proper \emph{recurrent neural
  network} (RNN), trained with the said observed time series, provides
well-timed estimations of future data.
The ensamble of the said wavelet analysis and RNN is called
WRNN~\cite{capizzi12a,napoli10b,napoli10a} and provides forecasts for
the number of users that will share a fragment.  Several neural
networks have been employed to find polaritons propagation and metal
thickness correspondence~\cite{Bonanno14}; to predict the behaviour of
users requesting resources~\cite{napoli13}, to perform wavelet
transform in a recursive lifting
procedure~\cite{capizzi12b,Sweldens98}.

\begin{figure}
  \centering
  \includegraphics[width=.36\textwidth]{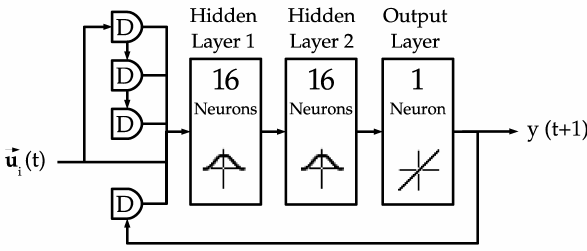}
  \caption{Topology of the WRNN}
  \label{f:nnet2}
\end{figure}

\subsection{WRNN setup}
For this work, the initial datasets consists of a time series
representing requests for a torrent, coming from peers, or given
declaration of
 availability for the torrent coming from both peers and seeds.
Independently of the specificities of such data, let us call this
series $x(\tau)$, where $\tau$ is the discrete time step of data,
sampled with intervals of one hour.  A biorthogonal wavelet
decomposition of the time series has been computed to obtain the
correct input set for the WRNN as required by the devised architecture
shown in Figure~\ref{f:nnet2}.  This decomposition can be achieved by
applying the wavelet transform as a recursive couple of conjugate
filters in such a way that the $i$-esime recursion $\hat{W}_i$
produces, for any time step of the series, a set of coefficients $d_i$
and residuals $a_i$, and so that
\begin{equation}
\hat{W}_i [a_{i-1}(\tau)] = [d_i(\tau) ,  a_i(\tau) ] ~~~~~ \forall~ i \in [1,M]\cap\mathbb{N}
\label{eq:xtau}
\end{equation}
where we intend $a_0(\tau)=x(\tau)$. The input set can  then be
represented as an $N \times (M+1)$ matrix of 
$N$ time steps of a $M$ level wavelet decomposition,
where the 
$\tau$-esime row represents the $\tau$-esime time step as the
decomposition 
\begin{equation}
\mathbf{u}(\tau) = \left [ d_1(\tau), d_2(\tau) , \ldots , d_M(\tau)
,a_M(\tau) \right ] 
\label{eq:utau}
\end{equation}

Each row of this dataset is given as input value to the $M$ input
neurons of the proposed WRNN (Figure~\ref{f:nnet2}).  The properties
of this network make it possible, starting from an input at a time
step $\tau_n$, to predict the effective number of requests (or offers)
at a time step $\tau_{n+r}$.  In this way the WRNN acts like a
functional
\begin{equation}
\hat{N}[\mathbf{u}(\tau_n)] = \tilde{x}(\tau_{n+r})
\label{eq:Nfunctional}
\end{equation}
where $r$ is the number of time steps of forecast in the future and
$\tilde{x}$ the predicted serie.

\subsection{Predicted user behaviour}

As described by equation~\eqref{eq:Nfunctional}, it is then possible
to obtain a future prediction of
the number of requests for a specific torrent,  as well as its
availability in the future.  In fact, by  considering both the
predicted $\tilde{x}_k(\tau_{n+r})$ and the modeled
$\chi_k(\tau_{n+r})$, it is possible, at a time step $\tau_n$, to
take counteracting actions 
and improve the probability of diffusion
for a rare torrent.
This is achieved, in practice, by using altered values for
$D_k(\tau_{n+r})$, which account for the forecast of future time
steps.
Such modified values are computed by our WRNN, e.g.\ in a computing node of
a cloud.  Therefore, predicted values for $T_k(\tau_{n+r})$,
$S_k(\tau_{n+r})$ and $\rho_k(\tau_{n+r})$ are sent to each node
acting as a peer.

Each time a new torrent becomes shared on the P2P network, then a new
WRNN is created and trained on a server, e.g.\ requested from a cloud
system,  to provide predictions related to the availability
and peers of the novel set of shared fragments for that torrent.  The
predictions will be sent to the peers periodically, and allow peers to
update the values of $D_k(\tau)$. The update frequency can be tuned in
order to correctly match the dynamic of hosts.


\begin{figure*}
  \centering
  \includegraphics[width=.24\textwidth]{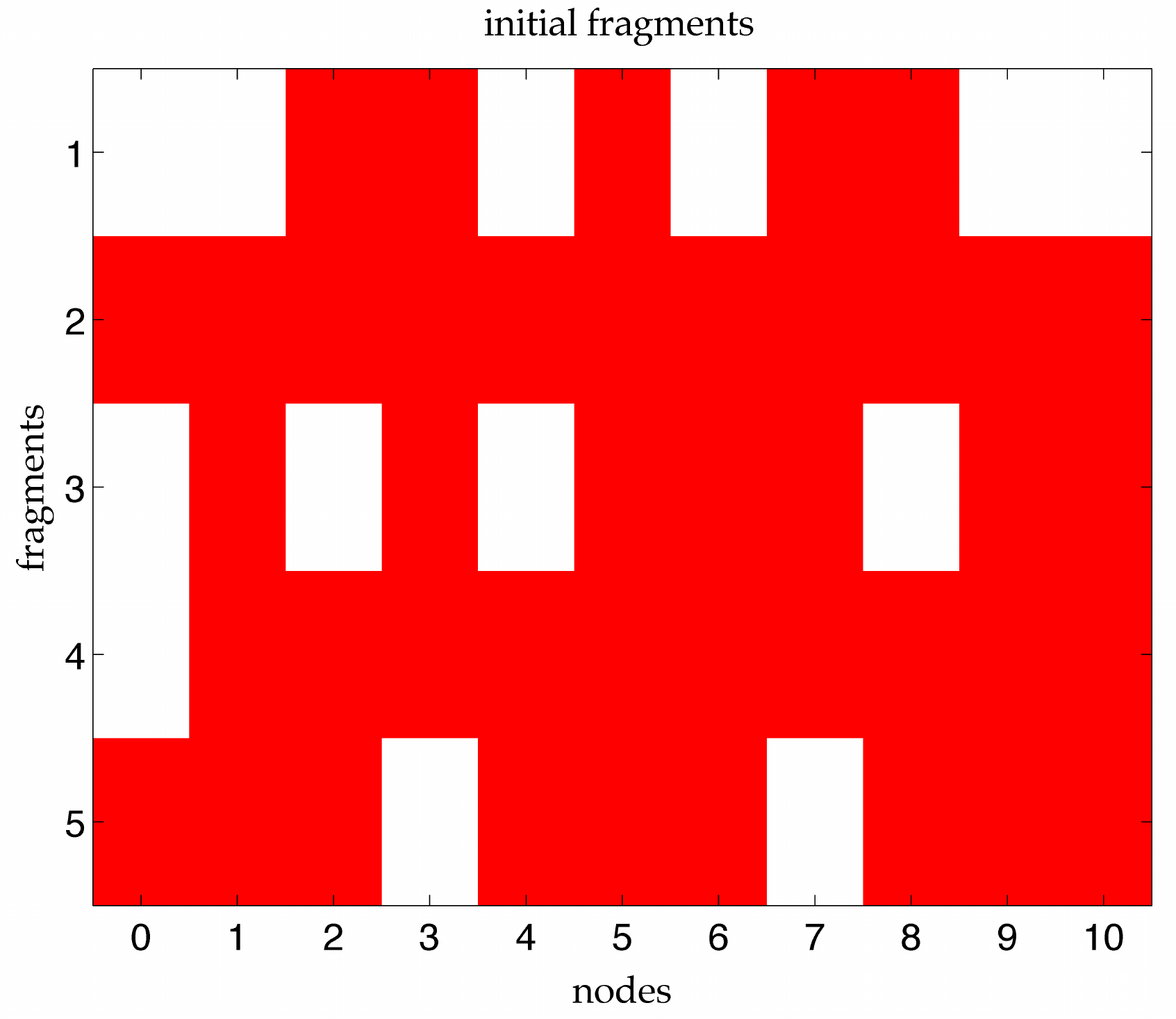}~
  \includegraphics[width=.24\textwidth]{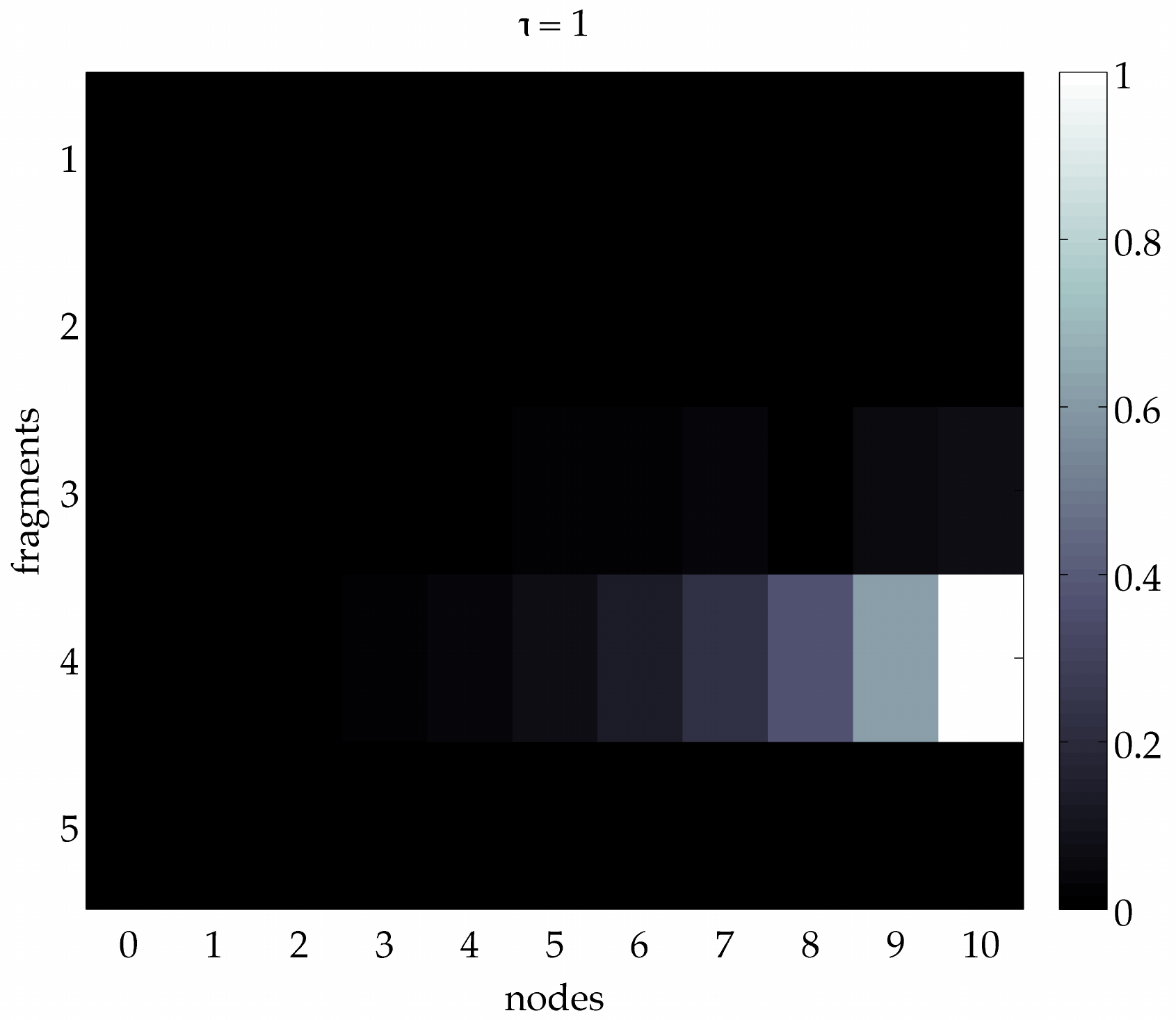}~
  \includegraphics[width=.24\textwidth]{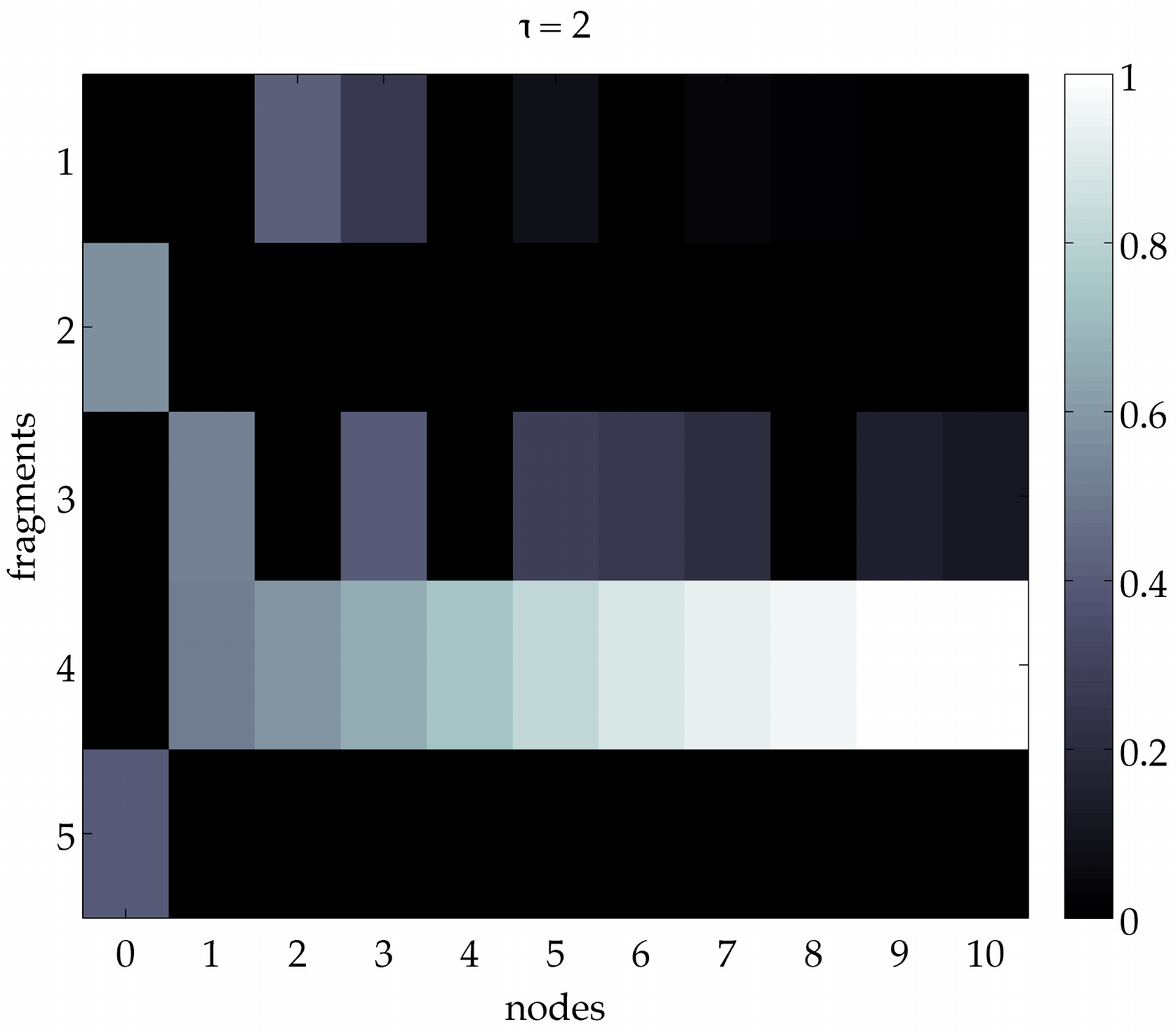}\\
  \includegraphics[width=.24\textwidth]{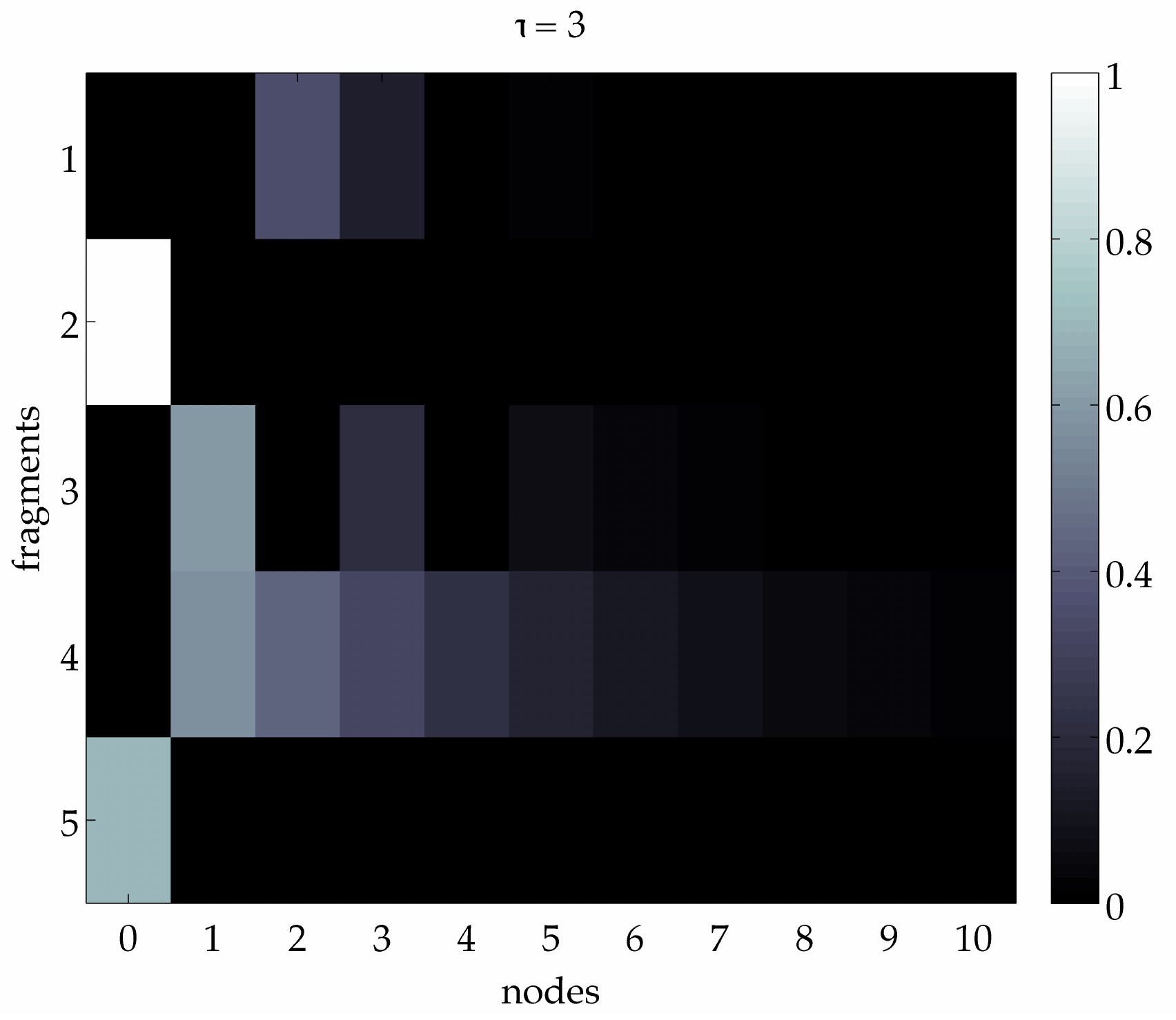}~
  \includegraphics[width=.24\textwidth]{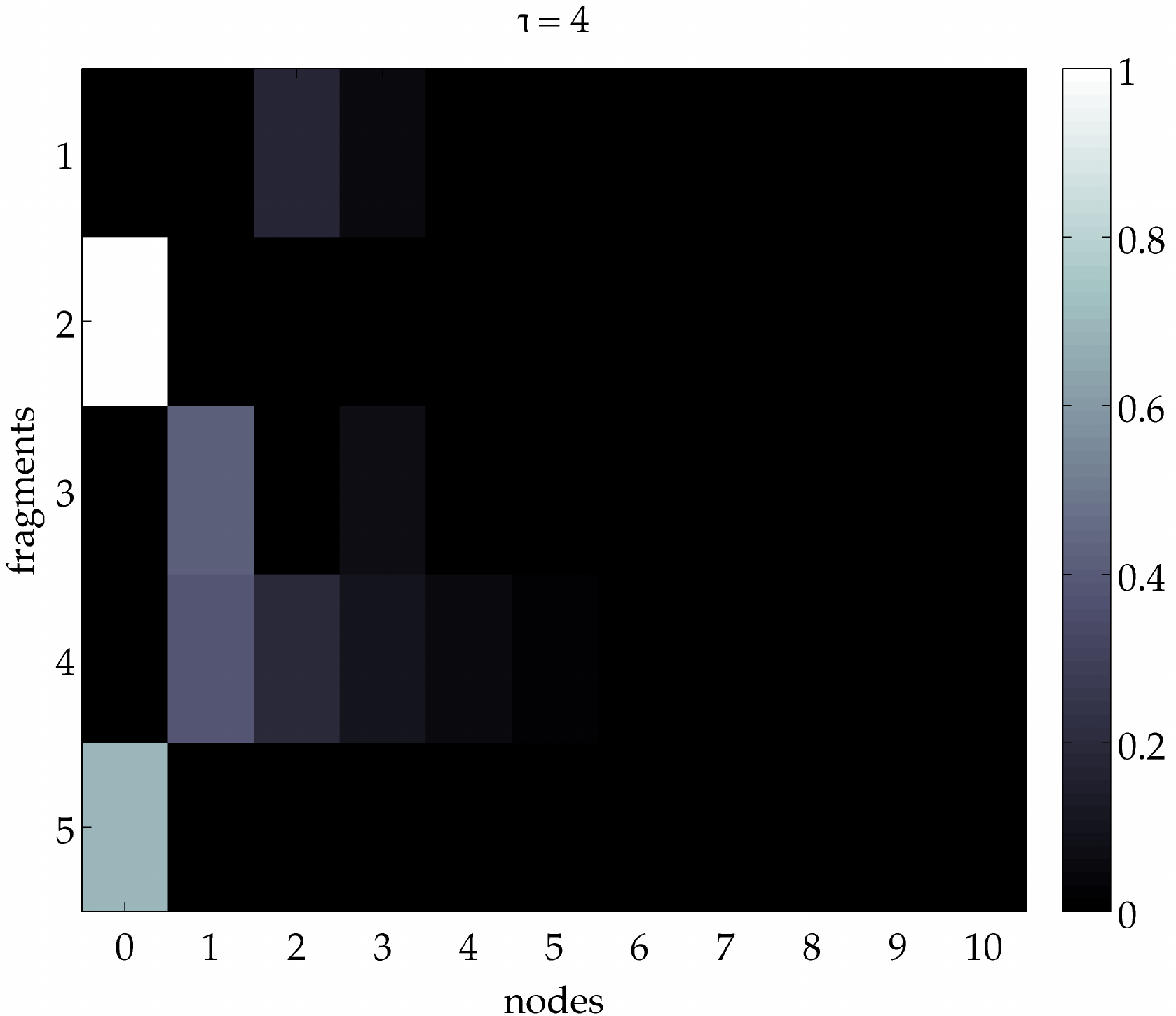}~
  \includegraphics[width=.24\textwidth]{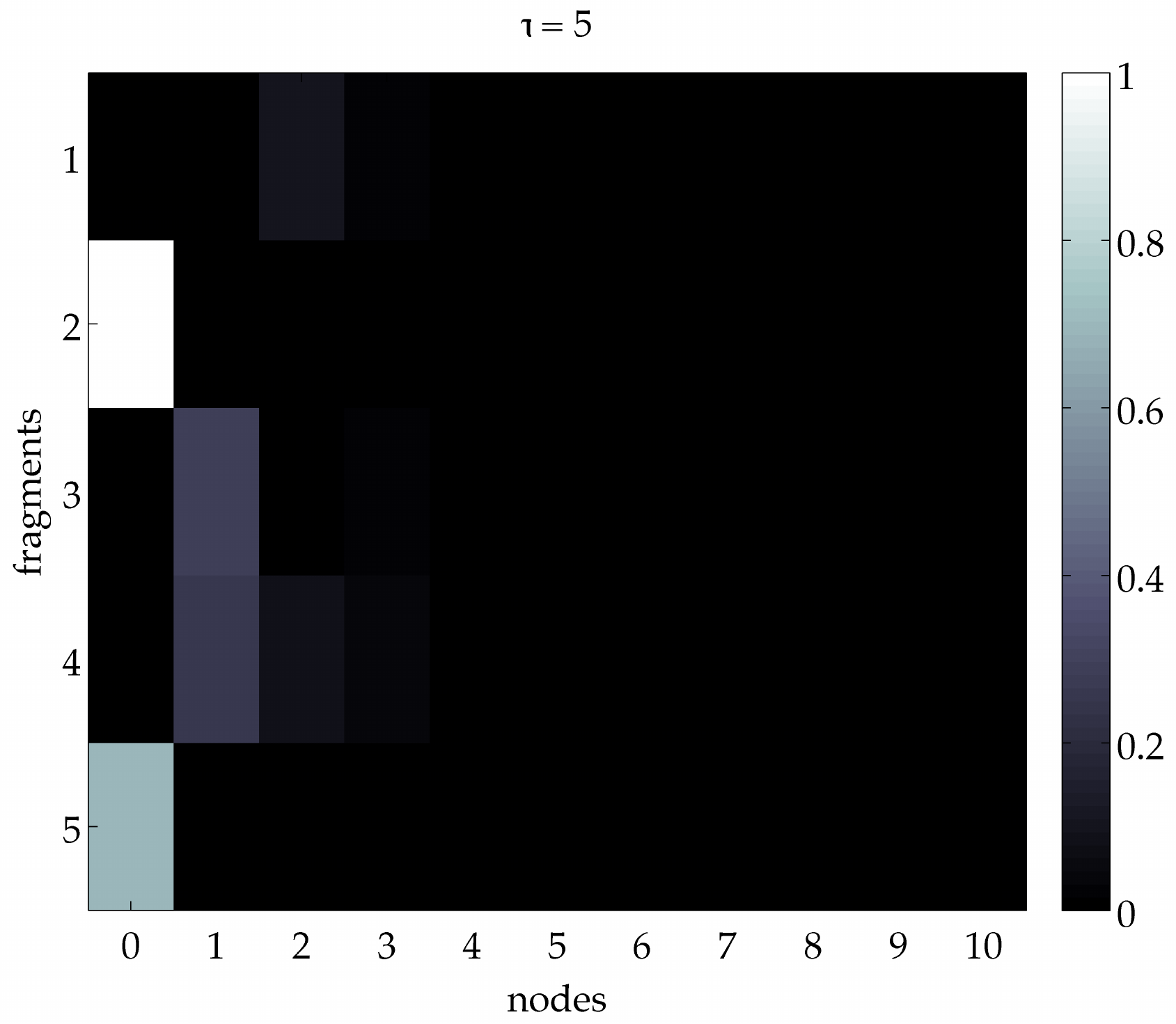}
  \caption{In order: the initial condition (white cells represent the
    fragments), the normalised $\chi^{ij}_k(\tau)$ for a certain node
    $n^i$ at different time steps $\tau$} 
  \label{f:sim}
\end{figure*}

\section{Experiments}
\label{experiments}

As shown in Figure~\ref{f:sim}, for the initial condition of the P2P
system some file fragments for a torrent happen to be heterogeneously
spread among peers (e.g.\ no one shares fragment n.\ 2, and very few
nodes have fragment n.\ 5).
We report a simulation comprising 11 peers and 5 file fragments, and
an evolution in only 5 time steps.  In the order, step after step,
peer $n^0$ selects file fragment $z_4$ and sends this to peer
$n^{10}$.  Both the fragment to spread and the destination peer
have been chosen according to equation~\eqref{eq:min}.

Later on, as soon as possible, peer $n^0$ selects another fragment,
i.e.\ $z_3$ to spread.  Such a fragment could be send just after the
transfer of the previous fragment has been completed, or concurrently
to the first transmission.
%
The shown evolution does not consider that the file fragment could
have been passed, e.g., to node $n^{10}$,
%
and so that for the next time step the value of $\chi$ for $n^{10}$
would drop to zero.  Note that the highest values of $\chi$ are an
indication of the urgency of receiving a fragment.

\begin{figure}
  \centering
  \includegraphics[width=.34\textwidth]{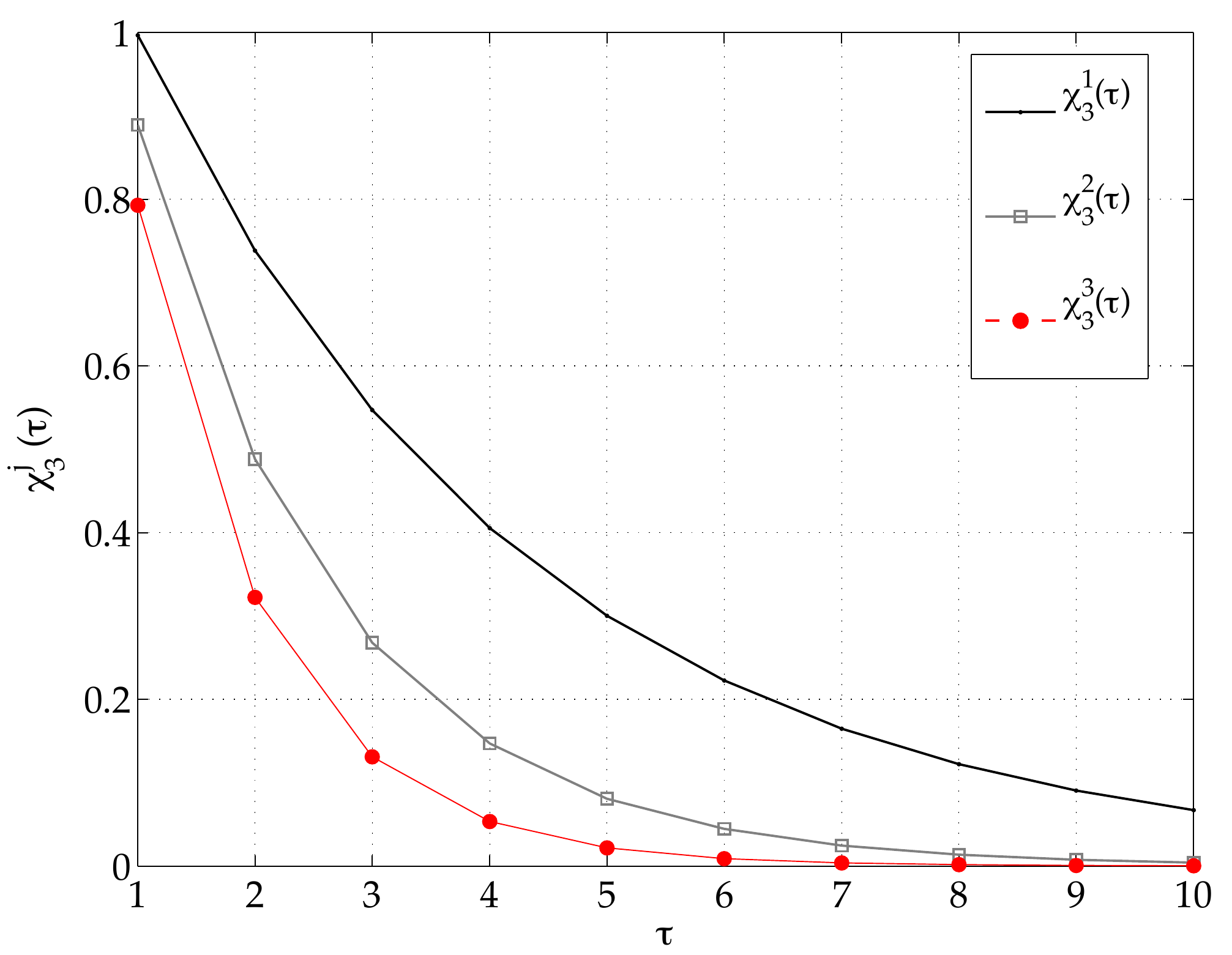}
  \caption{Time decay of normalised $\chi^{j}_k(\tau)$ for
    increasing time steps $\tau$}  
  \label{f:decay}
\end{figure}

The described model and formula allow subsequent sharing activities,
after the initial time steps, to be determined, in terms of which
fragments should be sent.
Figure~\ref{f:sim} shows that after the first time steps it becomes
more and more urgent for node $n^0$ to obtain the missing fragments
$z_2$ and $z_5$.
%
%
It is possible to see that the highest priority is for fragment $z_2$
since its share ratio and the relative availability are very low with
respect to fragment $z_5$.  This was the expected behaviour of the
developed model.
For the simulation shown in Figure~\ref{f:sim}, all fragments, except
$z_2$ since it is actually unavailable, would be spread to peers
 in a very low number of time steps.
%
%
Figure~\ref{f:decay} shows the decay of several computed $\chi$ for
different peers requiring fragment number 3.  On the long run, this
law will benefit nearby nodes, while on the short term, distant nodes
are given the highest priority.

%
%


\section{Related Work}
\label{related}

Several studies have analysed the behaviour of BitTorrent systems from
the point of view of fairness, i.e.\ how to have users contribute with
contents that can be uploaded by other users, levelling the amount of
downloads with that of upload. 
Fewer works have studied the problem of unavailability of contents in
P2P BitTorrent networks.
In~\cite{Qiu04}, authors proposed to order peers according to their
uploading bandwidth, hence when providing contents the selection of
peers is performed accordingly. 
One of the mechanism proposed to increase files availability has been
to use multi-torrent, i.e.\ for ensuring fairness, instead of forcing
users stay longer, they provide their contribution to uploaders for
fragments belonging to different files~\cite{Guo05}.
Similarly, in~\cite{Kaune10} authors show
that by using multi-torrent availability can be easily increased, and
confirm that fast
replication of rare fragments is essential.
Furthermore, 
bundling, i.e.\ the dissemination of a number of related files
together, has been proposed to increase
availability~\cite{Menasche09}.


The above proposed mechanisms differ from our proposal, since we take
into account several novel factors: the dynamic of data exchange
between distant peers, a decay for the availability of peers, and the
forecast of contents availability.
Such factors have been related into a proposed model that manages to
select the rarest content to be spread taking into account the future
availability, and the peers that should provide and take such a
content.

\section{Conclusions}
\label{conclusion}
This paper proposed to improve availability of fragments on a P2P
system by adopting a mathematical model  and a  neural network.  The
former describes the fragments diffusion and the urgency to share
fragments, whereas the latter provides an estimation of the
availability of peers, and hence fragments, at later time. 
By using the estimate of future availability into the diffusion model,
we can select the fragments that need to be spread to counteract their
disappearance due to users disconnections.

\section*{Acknowledgment}

This work has been supported by project PRISMA PON04a2 A/F funded by
the Italian Ministry of University and Research within PON 2007-2013
framework.

\bibliographystyle{abbrv} 

\bibliography{IEEEabrv,p2pbib}

\end{document}